\documentclass{iau} 
\usepackage{graphicx}

\title[X-rays from Radio Millisecond Pulsars] 
{X-rays from Radio Millisecond Pulsars}

\author[Slavko Bogdanov]   
{Slavko Bogdanov$^1$}

\affiliation{$^1$Columbia Astrophysics Laboratory, Columbia University, \\ 550 West 120th Street, New York, NY 10027, USA \\ email: {\tt slavko@astro.columbia.edu} }

\pubyear{2017}
\volume{337}  
\jname{Pulsar Astrophysics -­ The Next 50 Years}
\editors{P. Weltevrede, B.B.P. Perera, L. Levin Preston \& S. Sanidas, eds.}

\begin{document}

\maketitle
\begin{abstract}
The Galactic population of rotation-powered (aka radio) millisecond pulsars (MSPs) exhibits diverse X-ray properties. Energetic MSPs  show pulsed non-thermal radiation from their magnetospheres. Eclipsing binary MSPs predominantly have X-ray emission from a pulsar wind driven intra-binary shock. Typical radio MSPs emit X-rays from their heated magnetic polar caps. These thermally emitting MSPs offer the opportunity to place interesting constraints on the long sought after dense matter equation of state, making them important targets of investigation of the recently deployed Neutron Star Interior Composition Explorer (NICER) X-ray mission. 
\end{abstract}
\firstsection 
\section{Introduction}
\begin{figure}[t]
\begin{center}
 \includegraphics[angle=0,width=4.4in]{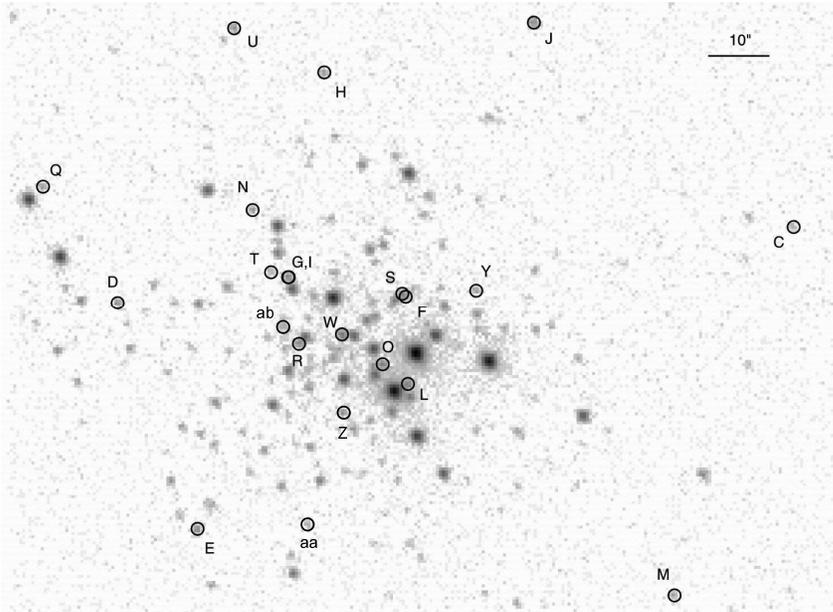} 
 \caption{A deep \textit{Chandra} ACIS-S 0.3--8 keV image of the core of the globular cluster 47 Tuc. The $1''$ radius circles are centered on the 22 known radio MSPs in this cluster, all of which have faint X-ray counterparts. See \cite{Bhattacharya17} for further details.}
   \label{fig1}
\end{center}
\end{figure}

Rotation-powered MSPs were first identified as X-ray sources with \textit{ROSAT} by \cite[Becker \& Tr\"umper~(1993)]{Becker93}. Subsequent studies with  \textit{ROSAT}, \textit{BeppoSAX}, \textit{RXTE}, and \textit{ASCA} in the 1990's were limited to a handful of the most luminous or nearest MSPs (see, e.g., \cite[Becker \& Tr\"umper~1999]{Becker99} and references therein). The launch of \textit{Chandra} and \textit{XMM-Newton} in 1999 enabled a quantum leap in our understanding of the X-ray emission properties of radio MSPs. A number of sensitive timing and spectroscopic studies of nearby MSPs in the field of the Galaxy have been carried out with  \textit{XMM-Newton} (e.g., \cite[Webb et al.~2004]{Webb04}; \cite[Kuiper et al.~2004]{Kuiper04}; \cite[Zavlin~2006]{Zavlin06}; \cite[Bogdanov \& Grindlay~2009]{Bogdanov09}; \cite[Bogdanov~2013]{Bogdanov13}; \cite[Spiewak et al.~2016]{Spiewak16}). The superb angular resolution of \textit{Chandra} has made possible  complementary studies of globular clusters, where MSPs are abundant and can be studied efficiently in large numbers via deep exposures. To date, X-rays have been detected from numerous radio MSPs in the globular clusters 47 Tuc (see Fig.~\ref{fig1} and \cite[Bhattacharya et al.~2017]{Bhattacharya17}), NGC 6397 (\cite[Bogdanov et al.~2010]{Bogdanov10}), M28 (\cite[Bogdanov et al.~2011a]{Bogdanov11a}), M4 (\cite[Bassa et al.~(2004)]{Bassa04}), M30 (\cite[Ransom et al.~2004]{Ransom04}), M71 (\cite[Elsner et al.~2008]{Elsner08}), Terzan 5 (\cite[Bogdanov~2008]{Bogdanov 2008}) and NGC 6752 (\cite[Forestell et al.~2014]{Forestell14}). These studies have established that radio MSPs tend to be relatively low luminosity X-ray emitters ($L_X \lesssim 10^{32}$ erg s$^{-1}$) in the soft X-ray band (0.1--10 keV) aside from three notable exceptions. The wealth of observational information collected over the past two decades has revealed that MSPs can be grouped into three broad categories based on the dominant X-ray production mechanism: i) non-thermal pulsed emission from the magnetosphere; ii) orbitally modulated emission from an intra-binary shock; iii) thermal emission from the magnetic polar caps. Herein, I provide a brief summary of the observational information gathered in X-rays on these categories of rotation-powered MSPs thus far and discuss future prospects.

\section{The X-ray Properties of Rotation-powered MSPs}
\subsection{Non-thermal Pulsed X-ray Emission}
\begin{figure}[t]
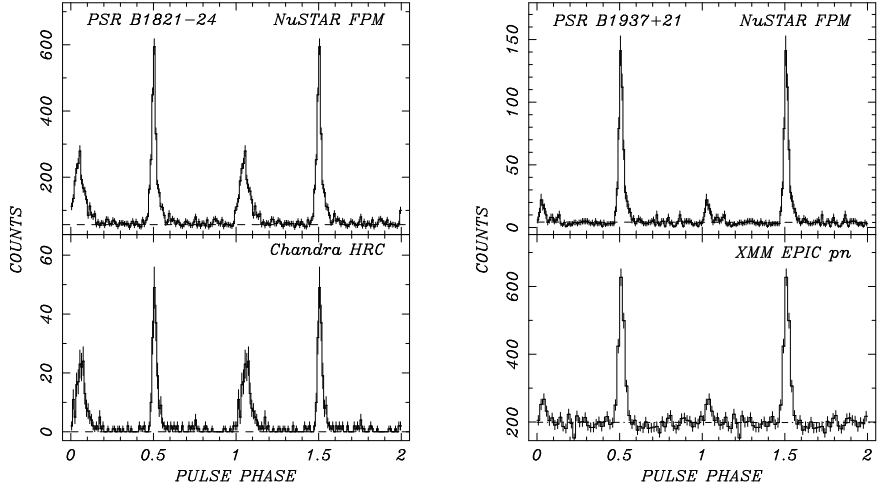

\begin{center}
 \includegraphics[angle=-90,width=2.1in]{fig2.ps}~~~~~~~
 \includegraphics[angle=-90,width=2.1in]{fig6.ps} 
 \caption{Non-thermal X-ray pulsations of the two most energetic radio MSPs, PSRs B1821$-$24 (\textit{left}) and B1937$+$21 (\textit{right}), in the hard (\textit{top}) and soft (\textit{bottom}) X-ray bands. Two rotational cycles are shown for clarity. Figure adapted from \cite[Gotthelf \& Bogdanov~(2017)]{Gotthelf17}.}
   \label{fig2}
\end{center}
 \end{figure}

Of the dozens of radio MSPs detected in X-rays to date, only three exhibit narrow pulses (Fig.~\ref{fig2}) with power-law spectra: PSRs B1821$-$24 , B1937$+$21, and J0218$+$4232 (see, e.g., \cite[Zavlin~2007]{Zavlin07} and references therein), which happen to be the most energetic, youngest, and with the highest magnetic fields at the light cylinder. The non-thermal pulsed X-rays are likely generated by the same process as in young ordinary pulsars, namely, relativistic particles accelerated in the pulsar magnetosphere.  Most recently, timing observations with \textit{NuSTAR} have shown that the same pulsed emission extends well into the hard X-ray band, up to at least $\sim$50 keV for PSR B1821$-$24 (\cite[Gotthelf \& Bogdanov~2017]{Gotthelf17}), with no evidence of  a spectral break or turnover. 

 \subsection{Intra-binary Shock Emission}
 High-energy radiation due to a shock driven by the interaction of the pulsar wind with material from a close companion star had been predicted for eclipsing binary MSP systems (e.g., \cite[Arons \& Tavani~1993]{Arons93}) a decade before it was conclusively observed (\cite[Stappers et al.~2003]{Stappers03}; \cite[Bogdanov et al.~2005]{Bogdanov05}). Since then, it has been determined that some ``black widow'' and  all ``redback'' MSP binaries show hard non-thermal X-ray emission with  $\approx$$10^{31-32}$ erg s$^{-1}$ (0.1--10 keV) and characteristic orbital phase dependence of the flux (e.g., \cite[Bogdanov et al.~2011b]{Bogdanov11b}; \cite[Bogdanov et al.~2014]{Bogdanov14}; \cite[Gentile et al.~2014]{Gentile14}; \cite[Tendulkar et al.~2014]{Tendulkar14};  see also contribution by M.~S.~E.~Roberts in this volume).  Detailed modeling of this shock emission can, in principle, serve as a powerful diagnostic of rotaion-powered MSP winds and collisionless shocks (\cite[Romani \& Sanchez~2016]{Romani16}; \cite[Wadiasingh et al.~2017]{Wadiasingh17}).  
 
 \subsection{Surface Thermal X-ray Emission}
\begin{figure}[t]
\begin{center}
  \includegraphics[height=2.65in]{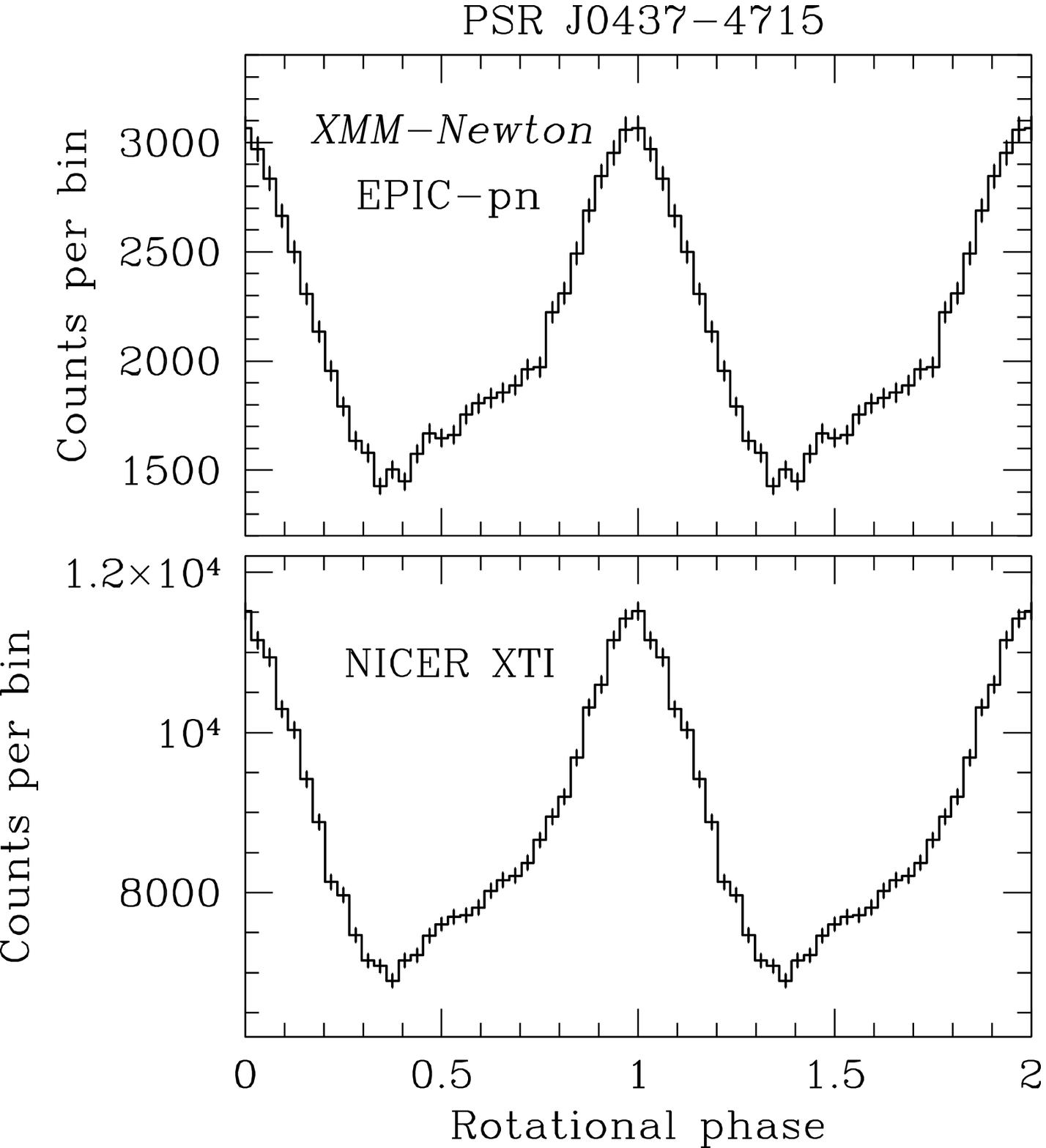}~~~~~
 \includegraphics[height=2.65in]{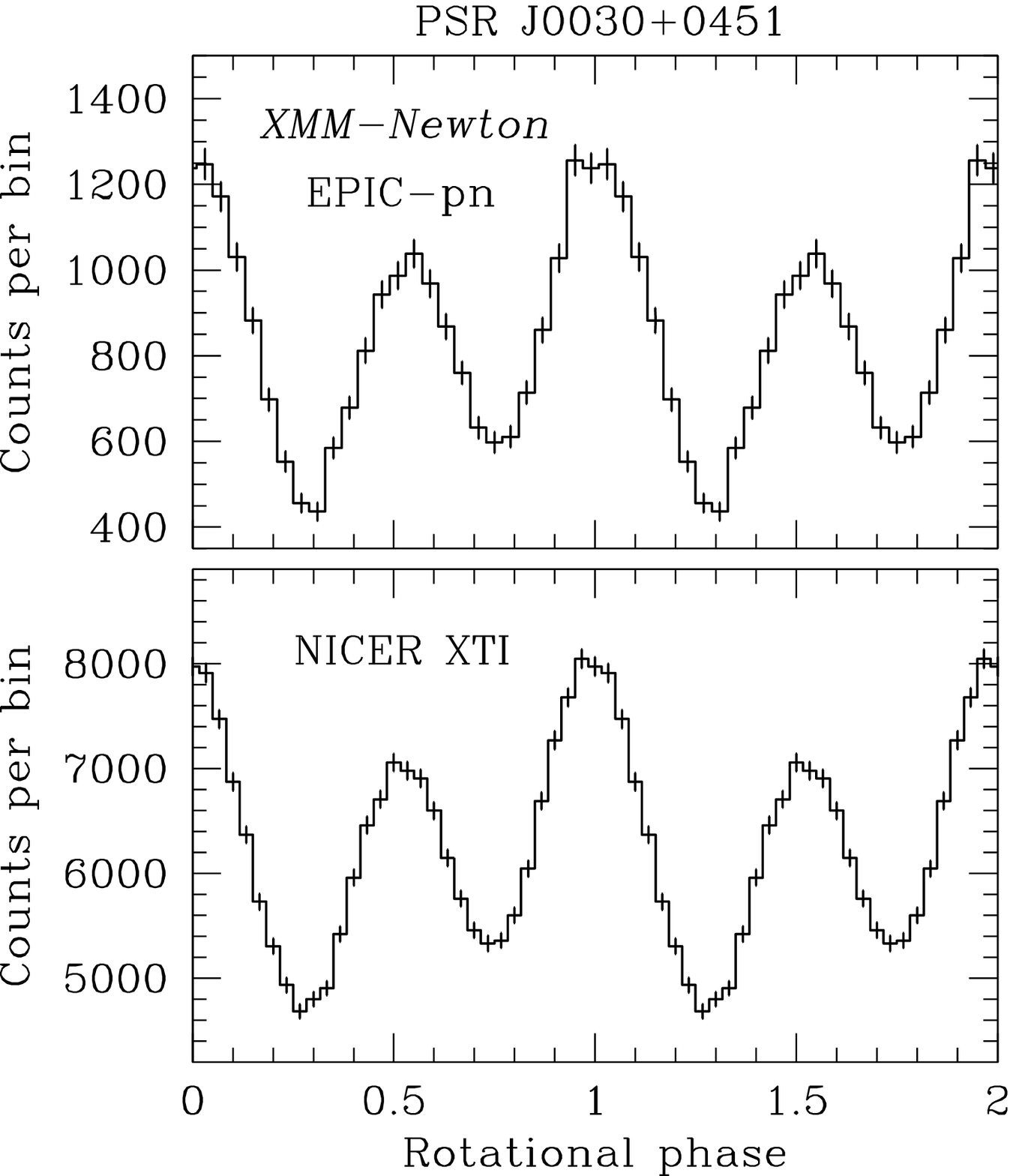}
 \caption{Thermal X-ray pulsations (0.3--2 keV) fromf two of the nearest MSPs, PSRs J0437$-$4715 (\textit{left}) and J0030$+$0451 (\textit{right}), from \textit{XMM-Newton} data (top)  and from a preliminary analysis of NICER XTI data (\textit{bottom}) presented here courtesy of the NICER team.}
   \label{fig3}
\end{center}
\end{figure}

The majority of known rotation-powered MSPs fall in the category of thermally-emitting X-ray sources. They exhibit broad X-ray pulsations (Fig.~\ref{fig3}) with soft blackbody-like spectra and X-ray luminosity $\le$$10^{31}$ erg s$^{-1}$ (e.g., \cite[Zavlin \& Pavlov~1998]{Zavlin98}; \cite[Zavlin 2006]{Zavlin06}), reaching down to $\sim$$10^{29}$ erg s$^{-1}$ (\cite[Pavlov et al.~2007]{Pavlov07}; \cite[Swiggum et al.~2017]{Swiggum17}). This emission appears to be localized in regions much smaller than the whole stellar surface and likely originates from the magnetic polar caps of the MSP, which are heated by a return flow of relativistic particles from the open field line region (see, e.g., \cite[Harding \& Muslimov~2002]{Harding02}).
Thermal emission from neutron stars is of great interest since photons emitted from the surface carry valuable information about the stellar compactness, i.e., the mass and radius. By measuring the mass-radius relation of several neutron stars to  $\lesssim$10\% it is possible to set strong constraints on the state of cold matter at densities exceeding those of atomic nuclei (e.g., \cite[Hebeler et al.~2013]{Hebeler13}). As part of its core science program, NASA's NICER  X-ray timing mission (see \cite[Arzoumanian et al.~2014]{Arzoumanian14} and contribution by P.~S.~Ray in this volume), which was deployed in June of 2017,  will target several of the nearest thermally emitting rotation-powered MSPs in very deep exposures.  Applying sophisticated models, which take into account all practically important general and special relativistic effects, rotation-induced oblateness of the star, neutron star atmospheric emission properties, and geometric configuration, to the X-ray pulse profiles of these radio MSPs is expected to produce a $\sim$5\% measurement of the neutron star radius (see, e.g., \cite[Miller \& Lamb~2015]{Miller15}; \cite[\"Ozel et al.~2016]{Ozel16}). 

\section{Future Prospects}
To date, all detections of pulsed X-ray emission from rotation-powered MSPs have relied on previous determination of the pulsar spin period at radio wavelengths. With the future generation of high-throughput X-ray observatories such as ATHENA and Lynx it will be possible to undertake the first systematic sensitive blind pulsation searches that may uncover many previously unknown nearby field and globular cluster rotation-powered MSPs.  This approach has the promise to yield a more complete census of MSPs, since it has recently come to light that some MSPs are undetectable in the radio due to faintness or perpetual eclipses.

\end{document}